\documentclass[aps,floatfix,prd,twocolumn,nofootinbib]{revtex4}  
\usepackage{graphicx}   
\usepackage{amsmath}
\usepackage{amssymb}
\usepackage{eepic,epic}

\newcommand{\beq}{\begin{equation}}
\newcommand{\eeq}{\end{equation}}
\newcommand{\dsl}{{\partial \hspace{-2.0mm}/}}
\newcommand{\Tr}{\mathrm{Tr}}
\newcommand{\SU}[1]{\mathrm{SU}\left( #1\right)}
\newcommand{\MeV}{\;\mathrm{MeV}}
\newcommand{\av}[1]{\langle {#1} \rangle}
\newcommand{\scalar}{{\left(s\right)}}
\newcommand{\pscalar}{{\left(p\right)}}
\newcommand{\cflpi}{{CFL$\pi^\pm$ }}
\newcommand{\cflkp}{{CFLK$^\pm$ }}
\newcommand{\cflk}{{CFLK$^0$ }}
\newcommand{\cflks}{{CFLK$^{0*}$ }}
\newcommand{\Ref}[1]{Ref.~\cite{#1}}
\newcommand{\Refs}[1]{Refs.~\cite{#1}}
\newcommand{\Eq}[1]{Eq.~(\ref{#1})}
\newcommand{\Eqs}[1]{Eqs.~(\ref{#1})}
\newcommand{\Fig}[1]{Fig.~\ref{#1}}
\newcommand{\Figs}[1]{Figs.~\ref{#1}}

\begin{document}   

\title{NJL model of homogeneous neutral quark matter:\\
       Pseudoscalar diquark condensates revisited}

\author{H.~Basler}
\affiliation{Institut f\"ur Kernphysik, Technische Universit\"at Darmstadt,
  Germany}

\author{M.~Buballa}
\affiliation{Institut f\"ur Kernphysik, Technische Universit\"at Darmstadt,
  Germany}

\begin{abstract}
We use a Nambu-Jona Lasinio type model to investigate the phase diagram of 
dense quark matter under neutron star conditions in mean field approximation. 
The model contains selfconsistently determined quark masses and allows
for diquark condensation in the scalar as well as in the pseudoscalar
channel. The latter gives rise to the possibility of $K^0$ condensation in 
the CFL phase. In agreement with earlier studies we find that this 
\cflk phase covers large regions of the phase diagram and that the 
predominant part of this phase is fully gapped. We show, however, that
there exists a region at very low temperatures where the \cflk solutions
become gapless, possibly indicating an instability towards anisotropic or
inhomogeneous phases. 
The physical significance of solutions with pseudoscalar diquark condensates
in the 2SC phase is discussed as well.
\end{abstract}

\date{\today}

\maketitle

\section{Introduction}
\label{sec:introduction}

Matter at high density and low temperature is expected to be a color
superconductor. (For reviews see, for example, \Refs{Rajagopal:2000wf,
  Alford:2001dt,Schafer:2003vz, Rischke:2003mt, Buballa:2003qv,
  Shovkovy:2004me, Alford:2007xm}).  At very high density, in the
limit where QCD is weakly coupled, it is possible to investigate color
superconductivity from first principles. From these studies
\cite{Schafer:1999fe,Evans:1999at} we know that the ground state of
three-flavor matter at asymptotically high density is the Color-Flavor
Locked (CFL) phase~\cite{Alford:1998mk}, where quarks of all colors
and flavors are paired in a homogeneous scalar ($J^P = 0^+$) diquark
condensate.  Unfortunately, these analyses cannot directly be applied
to describe the core of a neutron star, which is the most likely place
for color superconducting quark matter to exist in nature.  Although
densities of several times nuclear matter density can be reached
there, this is far from being asymptotic, and the weak-coupling
expansion breaks down.

A second difficulty arises from the fact that, at these ``moderate''
densities, the strange quark mass cannot be neglected against the
chemical potential.  Moreover, in compact stars the system must be in
beta equilibrium and electrically and color neutral.  As a
consequence, the Fermi momenta of the different quark flavors are
unequal in the unpaired system, and their BCS pairing is only favored
if the free-energy cost to adjust the Fermi surfaces is
overcompensated by the pairing energy.  It is therefore unclear
whether quark matter under compact star conditions is color
superconducting and, if so, which pairing pattern is most favored.
Besides the CFL phase, the most prominent candidate is the 2SC phase
\cite{Alford:1997zt, Rapp:1997zu}, where only up and down quarks are
paired.

The stress imposed by the different Fermi momenta of the quarks in a
diquark pair reduces the gap in the excitation spectrum.  Eventually,
the gap vanishes completely \cite{Sarma}, and one obtains a ``gapless
color superconducting state'', e.g., the gapless 2SC (g2SC) phase
\cite{Shovkovy:2003uu,Huang:2003xd} or the gapless CFL (gCFL) phase
\cite{Alford:2003fq,Alford:2004hz}.  However, at least at low
temperatures, these gapless phases suffer from chromomagnetic
instabilities \cite{Huang:2004bg,Casalbuoni:2004tb,Fukushima:2005cm},
indicating that they cannot be the true ground state of the system.
Possible alternatives are inhomogeneous phases, like mixed phases
\cite{Bedaque:1999nu,Neumann:2002jm,Reddy:2004my} or crystalline color
superconducting phases
\cite{Alford:2000ze,Bowers:2002xr,Casalbuoni:2003wh,GiannakisRen,
  Casalbuoni:2005zp,Mannarelli:2006fy,Rajagopal:2006ig,Nickel:2008ng,
  Sedrakian:2009kb}.

Another way the CFL phase can react on the stress induced by the
strange quark mass is to develop a K$^0$ condensate. The resulting
phase is called \cflk phase.  Since the K$^0$ has strangeness $+1$
(opposite to a strange quark, which has strangeness $-1$), its
condensation effectively reduces the strange quark content of the
system.  The K$^0$ condensation was predicted first within an
effective field theory approach based on the symmetry breaking pattern
of the CFL phase \cite{Schafer:2000ew,Bedaque:2001je,Kaplan:2001qk}:
The CFL condensates break the original $\SU{3}_{\mathrm{color}} \times
\SU{3}_L \times \SU{3}_R$ symmetry of QCD down to a residual
$\SU{3}_{\mathrm{color}+V}$ symmetry.  As a consequence of the
spontaneous breaking of chiral symmetry, there is an octet of
pseudoscalar Goldstone bosons, which are massless in the chiral limit.
Turning on finite quark masses, the Goldstone bosons remain rather
light \cite{Son:1999cm} whereas the strange quark mass acts as an
effective strangeness chemical potential, which eventually leads to
Bose condensation of the kaonic modes.

Since the kaon condensate reduces the stress in the diquark pairs, it
is expected to shift the occurrence of gapless modes to lower chemical
potentials. Within effective theories it was predicted that the first
mode which becomes gapless is electrically
charged~\cite{Kryjevski:2004jw, Kryjevski:2004kt}. In
\Ref{Gerhold:2006np} it was shown that all Meissner masses stay real
in this phase, but nevertheless a weak instability towards the
spontaneous generation of Goldstone-boson currents ($p$-wave kaon
condensates) was found~\cite{Kryjevski:2005qq,Schafer:2005ym}. At
lower chemical potential a neutral mode could become gapless as well,
leading to imaginary Meissner masses and stronger
instabilities~\cite{Gerhold:2006np}.

In order to decide, which of these various phases are realized under
compact star conditions, detailed calculations are
necessary. Unfortunately, as mentioned in the beginning, QCD
calculations based on a weak-coupling expansion are not applicable at
moderate densities.  In this situation the most promising development
is probably the Dyson-Schwinger approach, which has been successfully
extended to describe color superconducting quark matter
\cite{Nickel:2006vf,Nickel:2006kc,Marhauser:2006hy,Nickel:2008ef}.
However, because of the complexity of this method it is difficult to
apply it to the calculation of equations of state or phase diagrams.
So far, the most complete investigations of quark matter under compact
star conditions have therefore been performed within models.  Although
model calculations have only limited predictive power, they can deal
as explorative studies and may give valuable hints about possible
interesting scenarios or problems.  Among these models for color
superconducting quark matter, the most prominent ones are models of
the Nambu--Jona-Lasinio (NJL) type, because they allow for a
simultaneous and selfconsistent treatment of the various
condensates. Besides the diquark condensates this also includes the
chiral (quark-antiquark) condensates, which determine the dynamical
quark masses.

Phase diagrams for homogeneous electrically and color neutral quark
matter in NJL-type models have been calculated in
\Refs{Ruester:2005jc,Blaschke:2005uj, Abuki:2004zk,Abuki:2005ms}.
Both, diquark and quark antiquark condensates, were included, but the
authors neglected the possibility of Goldstone boson condensation.  A
rich phase structure was found, including large areas of gapless color
superconducting phases. Unfortunately, the details strongly depend on
the model parameters.

For the description of CFL$+$Goldstone boson condensed phases in
NJL-type models one has to take into account pseudoscalar diquark
condensates~\cite{Buballa:2004sx, Forbes:2004ww}.  In the chiral limit
the Goldstone bosons connect the continuous set of degenerate ground
states which are related to each other by axial $\SU{3}$
transformations, \beq q \rightarrow \exp \left(i\theta_a
  \frac{\tau_a}{2} \gamma_5 \right) q\,,
\label{eq:axtrans}
\eeq where $q$ is a quark field and $\tau_a$, $a = 1,\dots,8$, are the
Gell-Mann matrices in flavor space.  Under these transformations the
scalar diquark condensates of the CFL phase are partially rotated into
pseudoscalar ones.  In particular, the K$^0$ condensation is marked by
the formation of pseudoscalar condensates which correspond to
\Eq{eq:axtrans} with $a=6,7$, whereas charged pion or charged kaon
condensates correspond to $a=1,2$ or $a=4,5$, respectively.

In \Ref{Warringa:2006dk}, Warringa extended the NJL-model studies of
\Ref{Ruester:2005jc} in this direction.  Using the same model and the
same parameters but additionally allowing for pseudoscalar
condensates, he found that large parts of the CFL and gCFL phases are
replaced by the \cflk phase.  Moreover, Warringa reported that the
\cflk solutions contain no gapless modes, suggesting that no
instability problems occur.  On the other hand he found that, below a
certain threshold in the chemical potential, the electric charge
chemical potential at $T=0$ is nonzero in this phase.  This would mean
that there is a nonvanishing electron density, which is not allowed in
the fully gapped neutral \cflk phase at zero temperature.

The main motivation of our present study was to resolve this puzzle.
To that end we use the same model as in \Ref{Warringa:2006dk} and
carefully analyze the phase diagram and the excitation spectra.  As we
will see, to a large extent our results are in good agreement with
\Ref{Warringa:2006dk}.  The main difference is that we find a region
where the \cflk phase is gapless. However, this region is restricted
to extremely low temperatures, $T \ll 10$~keV.  We will also analyze
the so-called p2SC phase, a 2SC-like phase with pseudoscalar
condensates, which was found in \Ref{Warringa:2006dk} but has no
obvious physical interpretation.

The remaining part of this paper is organized as follows.  In
Sec.~\ref{sec:model} we introduce the model and briefly summarize how
the thermodynamic potential is obtained.  After that, in
Sec.~\ref{sec:results}, we discuss the resulting phase structure.  We
begin with a short overview in Sec.~\ref{ssec:phase_diagram} and then
perform a detailed analysis of the most interesting features.
Finally, in Sec.~\ref{sec:conclusions}, our results are summarized.

\section{Model}
\label{sec:model}

\subsection{Lagrangian}

We use the same NJL-type model as in \Refs{Ruester:2005jc,Warringa:2006dk}.
The Lagrangian reads
\beq
\mathcal{L} = \bar{q} (i\dsl -\hat{m} + \gamma_0\hat{\mu} ) q 
+ \mathcal{L}_{\bar{q}q} +
\mathcal{L}_{qq} +\mathcal{L}_6\,,
\label{L}
\eeq
where $q$ is a quark field with three color ($r$,$g$,$b$) 
and three flavor ($u$,$d$,$s$) degrees of freedom. $\hat{m}$ is the 
diagonal mass matrix $\hat{m} = \mathrm{diag}_f\left(m_u,m_d,m_s\right)$.

The diagonal matrix $\hat{\mu}$ contains the quark chemical potentials. 
As motivated in the Introduction, we are interested in the electrically 
and color neutral ground state in beta equilibrium. 
We then have four different conserved charges: the total quark 
number, the electric charge and two color charges. 
According to the corresponding charge densities $n$, $n_Q$, 
$n_3 = n_r-n_g$ and $n_8 = (n_r+n_g-2n_b)/\sqrt{3}$, we thus have 
four independent chemical potentials $\mu$, $\mu_3$, $\mu_8$, and $\mu_Q$. 
So $\hat{\mu}$ takes the form
\beq
    \hat{\mu} = \mu + \mu_Q Q + \mu_3 \lambda_3 + \mu_8 \lambda_8\,,
\eeq
with $Q = \mathrm{diag}_f \left(2/3,-1/3,-1/3\right)$ and the Gell-Mann
matrices $\lambda_3$ and $\lambda_8$ in color space.

The other terms in \Eq{L} describe the interaction.
\beq
\mathcal{L}_{\bar{q}q} = G \sum_{a=0}^8 \left[\left(\bar{q}\tau_a q\right)^2 +
  \left(\bar{q} i \gamma_5 \tau_a q \right)^2 \right]
\label{Lqbarq}
\eeq
is a four-point interaction in the quark-antiquark channel,
while
\begin{alignat}{1}
  \mathcal{L}_{qq} = H \hspace{-2mm} \sum_{A,B = 2,5,7} & \big[
  \left(\bar{q} i \gamma_5 \tau_A \lambda_{B} C \bar{q}^T\right)\left(
    q^T C i \gamma_5 \tau_A \lambda_B q \right)
\nonumber\\
+\; & \left(\bar{q} \tau_A \lambda_B C
    \bar{q}^T\right) \left( q^T C \tau_A \lambda_B q \right) \big]\,,
\label{Lqq}
\end{alignat}
is a four-point interaction in the quark-quark channel. 
$G$ and $H$ are dimensionful coupling constants.
In our notation $\lambda_a$ and $\tau_a$ are the Gell-Mann matrices 
acting in color space and in flavor space, respectively, where  
$\tau_0=\sqrt{2/3} \; {1\hspace{-1.4mm}1}_f$.
$C$ denotes the charge conjugation matrix $C = i \gamma^2 \gamma^0$.

Both, $\mathcal{L}_{\bar{q}q}$ and $\mathcal{L}_{qq}$, consist of a scalar
part (first term on the r.h.s.) and a pseudoscalar part (second term
on the r.h.s.) and are invariant under flavor $U(3)_L \times U(3)_R$ 
transformations. 
In order to break the $U(1)$ axial symmetry, we include a six-point
('t Hooft) interaction term~\cite{'t Hooft:1976fv}
\beq
\mathcal{L}_6 = -K \;\left\{\mathrm{det}_f
\left[\bar{q}\left(1+\gamma_5\right) q\right] + \mathrm{det}_f
\left[\bar{q} \left(1-\gamma_5\right) q \right]\right\}\,, 
\eeq
with coupling constant $K$.

\subsection{Phases and condensates}
\label{ssec:phases}

We introduce the following 
scalar and pseudoscalar diquark condensates\footnote{For the
  comparison with \Ref{Warringa:2006dk} it is important to note that we use
  the notation of \Ref{Buballa:2004sx} for the diquark condensates,
  $\Delta_\mathit{flavor,\;color}$, 
  while in~\Ref{Warringa:2006dk} they are denoted
  in the opposite way, $\Delta_\mathit{color,\;flavor}$.}
\begin{align}
\Delta_{AB}^{\scalar}  &= -2H\av{q^T C \gamma_5 \tau_A \lambda_B q}\,,
\\
\Delta_{AB}^{\pscalar} &= -2H\av{q^T C  \tau_A \lambda_B q}\,.
\end{align}
\begin{table}
  \begin{tabular}{|l|ccccccccc|}
    \hline
    &$\Delta_{22}^\scalar$ & $\Delta_{55}^\scalar$ & $\Delta_{77}^\scalar$ &
    $\Delta_{25}^\pscalar$ & $\Delta_{52}^\pscalar$ & $\Delta_{27}^\pscalar$ &
    $\Delta_{72}^\pscalar$ & $\Delta_{57}^\pscalar$ & $\Delta_{75}^\pscalar$\\  
    \hline
    NQ&-&-&-&-&-&-&-&-&-\\
    2SC&$\times$&-&-&-&-&-&-&-&-\\
    uSC&$\times$&$\times$&-&-&-&-&-&-&-\\
    p2SC&$\times$&-&-&$\times$&-&-&-&-&-\\
    CFL&$\times$&$\times$&$\times$&-&-&-&-&-&-\\
    \cflk&$\times$&$\times$&$\times$&$\times$&$\times$&-&-&-&-\\
    \cflkp&$\times$&$\times$&$\times$&-&-&$\times$&$\times$&-&-\\
    \cflpi&$\times$&$\times$&$\times$&-&-&-&-&$\times$&$\times$\\
    \hline
  \end{tabular}
  \caption{The phases investigated in this article. $\times$
    marks the presence of the condensates while $-$ means that the
    condensate vanishes. 
}
  \label{phases}
\end{table}

Depending on which of these diquark condensates take nonzero values, we can 
distinguish various color superconducting phases. The phases which have
been investigated in this article are summarized in Table~\ref{phases}. 
For instance, in the simplest possible color superconducting phase, 
the 2SC phase, only the $\Delta_{22}^{\scalar}$ condensate is present. 
In this phase only red and green up and down quarks form Cooper pairs 
while the strange and all blue quarks remain unpaired
(see Table~\ref{pairs}). In the uSC phase \cite{Iida:2004cj}
$\Delta_{22}^{\scalar}$ and $\Delta_{55}^{\scalar}$ are nonzero. 
Here red and blue up and strange quarks are paired, in addition to the
pairs which are already present in the 2SC phase. 

The most symmetric phase is the CFL phase, where 
$\Delta_{22}^\scalar$, $\Delta_{55}^\scalar$ and $\Delta_{77}^\scalar$ 
are non-zero, so that all quark flavors and colors are paired in a 
scalar condensate.
As in QCD, the presence of these condensates breaks the original 
$\SU{3}_\text{color} \times \SU{3}_\text{L} \times \SU{3}_\text{R}$ 
of the model  down to $\SU{3}_{\text{color} + \text{V}}$.
Among others, this gives rise to an octet of pseudoscalar Goldstone
bosons (see \Refs{Ebert:2006tc,Ebert:2007bp,Kleinhaus:2007ve,Kleinhaus:2008cd} 
for the NJL-model description of these Goldstone bosons). 
The condensation of some of these Goldstone bosons is then marked by
the appearance of nonvanishing pseudoscalar condensates. 
For instance, applying the transformation of \Eq{eq:axtrans} in the 
K$^0$ channel ($a = 6,7$) on the scalar condensates of the CFL phase
gives rise to the pseudoscalar condensates $\Delta_{25}^\pscalar$ and
$\Delta_{52}^\pscalar$. Similarly, in the \cflkp phase the condensates
$\Delta_{27}^\pscalar$ and $\Delta_{72}^\pscalar$, and in the \cflpi
phase the condensates $\Delta_{57}^\pscalar$ and
$\Delta_{75}^\pscalar$ appear.

In the 2SC phase, there are no pseudoscalar Goldstone bosons. 
Nevertheless, it was found in \Ref{Warringa:2006dk} that in some region 
of the phase diagram a so-called p2SC phase is favored, where a
pseudoscalar condensate is present in addition to $\Delta_{22}^{\scalar}$.
Therefore we will consider this possibility as well. 

On the other hand, we will not consider other combinations of scalar
diquark condensates, i.e., the dSC phase ($\Delta_{22}^\scalar \neq 0$ and 
$\Delta_{77}^\scalar \neq0 $), the sSC phase
($\Delta_{55}^\scalar \neq 0$ and $\Delta_{77}^\scalar \neq 0$),
the 2SC$_{us}$ phase ($\Delta_{55}^\scalar \neq 0$), or the
the 2SC$_{ds}$ phase ($\Delta_{77}^\scalar \neq 0$).
These possibilities were taken into account in 
Refs.~\cite{Ruester:2005jc} and \cite{Warringa:2006dk}, but were not
found to be realized in the phase diagram. So we do not expect them
to play a role.

In addition to the diquark condensates, we also introduce scalar 
antiquark-quark condensates
\beq
\phi_f = \av{\bar{q}_fq_f}\,, \quad f \in \{u,d,s\}\,,
\eeq
which are responsible for spontaneous chiral symmetry breaking at low 
temperatures and densities. 
Since chiral symmetry is explicitly broken by the mass matrix $\hat m$, 
this does not lead to new phases in a strict sense. However, it gives
rise to dynamical ``constituent'' quark masses (see \Eq{eq:Ma} below) 
which can be considerably larger than the bare masses contained in 
$\hat m$.

Expecting that the condensation of pions and kaons outside the
color superconducting phase does not occur, we neglect condensates of 
the form $\av{\bar{q} i \gamma_5 \tau_a q}$.
In principle, such condensates should be present in the 
CFL$+$Goldstone boson phases where they break no additional symmetry.
However, following \Ref{Warringa:2006dk}, we neglect them there as well,
expecting that their effect is small. Eventually, this should be checked.

\begin{figure}
\includegraphics[width=0.47\textwidth]{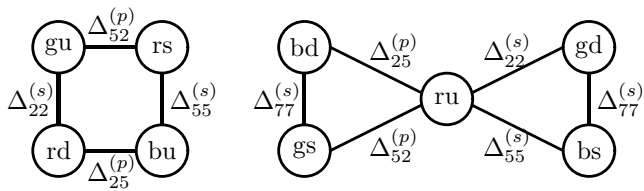}
\caption{Pairing pattern in the \cflk phase.}
\label{pairing}
\end{figure}

At the end of this section, we would like to add a few more details
about the \cflk phase and compare them with the more familiar CFL phase. 
Later on, this will be useful for the interpretation of the results.
In \Fig{pairing} we show the color-flavor structure of the \cflk pairing.
This pattern follows directly from Tables~\ref{phases} and \ref{pairs}.
If we switch off the pseudoscalar condensates, we arrive at the
CFL pairing pattern.
There the quarks can be grouped into four subsets, which are not
connected to each other by the pairing.
In the \cflk phase, the pseudoscalar condensates give rise to additional
connections, so that only two unconnected subsets remain. 
The smaller one consists of four pairs, connecting four different
quark species, the larger one contains six pairs, which connect
five quark species.

In Table~\ref{pairs} we have also listed the eigenvalues of the
``rotated'' electric charges
\beq
\tilde{Q} = Q - \frac{1}{2} \lambda_3 - \frac{1}{2\sqrt{3}} \lambda_8
\label{eq:qtilde}
\eeq 
for both constituents of each pair.
For the CFL phase it is well known that the ground state is
$\tilde Q$ neutral~\cite{Alford:1998mk,Alford:1999pb}. 
This follows from the fact that all quarks are paired and in each pair 
the sum $\tilde Q_1 + \tilde Q_2$ vanishes. 
Obviously, the same is true for the \cflk phase. 
An important consequence is that, at $T=0$, in both phases color neutral 
quark matter is automatically electrically neutral. Hence, no electrons
are admitted in neutral matter, i.e., 
$\mu_Q = 0$~\cite{Rajagopal:2000ff,Steiner:2002gx}.

%

\begin{table}[h]
  \begin{tabular}{|cccr|}
    \hline
    gap parameter& diquark pair & \quad $\tilde Q_1$ & $\tilde Q_2$ \quad\\  
    \hline
    $\Delta_{22}^\scalar $ & $ru-gd$ & \quad $0$ & $0$ \quad\\
                         & $gu-rd$ & \quad $1$ & $-1$ \quad\\
    \hline
    $\Delta_{55}^\scalar $ & $ru-bs$ & \quad $0$ & $0$ \quad\\
                         & $bu-rs$ & \quad $1$ & $-1$ \quad\\
    \hline
    $\Delta_{77}^\scalar $ & $gd-bs$ & \quad $0$ & $0$ \quad\\
                         & $bd-gs$ & \quad $0$ & $0$ \quad\\
    \hline
    $\Delta_{25}^\pscalar $ & $ru-bd$ & \quad $0$ & $0$ \quad\\
                         & $bu-rd$ & \quad $1$ & $-1$ \quad\\
    \hline
    $\Delta_{52}^\pscalar $ & $ru-gs$ & \quad $0$ & $0$ \quad\\
                         & $gu-rs$ & \quad $1$ & $-1$ \quad\\
    \hline
  \end{tabular}
  \caption{Gap parameters, color-flavor structure of the corresponding
           diquark pairs and $\tilde Q$ charges of the constituents. 
           The list is restricted to those condensates which are realized 
           in the phase diagrams of Sect.~\ref{sec:results}.        
                }
  \label{pairs}
\end{table}

While all pairs listed in Table~\ref{pairs} are $\tilde Q$ neutral
in total, we can distinguish two cases: either both quarks carry 
zero $\tilde Q$ charge (like in the $ru-gd$ pair) or the quarks
carry the $\tilde Q$ charges $\pm 1$ (like in the $gu-rd$ pair). 
Comparing this with \Fig{pairing}, we see that the two unconnected groups 
in the pairing pattern of the \cflk phase just correspond to these 
two cases.

\subsection{Thermodynamic potential}
\label{ssec:Omega}

In order to decide, which phase is favored at given temperature
and chemical potential, we calculate the thermodynamic 
potential per volume, 
\beq
\Omega = \frac{T}{V} \ln{\cal Z}\,, 
\eeq
where ${\cal Z}$ is the grand partition function.
Besides the quarks we consider the presence of noninteracting
leptons,
\beq
    \Omega = \Omega_q + \Omega_l\,,
\eeq
where $\Omega_l = \Omega_e + \Omega_\mu$ is the sum of the ideal gas
contributions of electrons and muons.
For the quark part $\Omega_q$ we restrict ourselves to the mean-field 
approximation and proceed in the same way as described in 
\Refs{Ruester:2005jc,Ruester:2006yh,Warringa:2006dk}.  
Therefore, we only summarize the main steps here
and refer to these papers for further details.

Linearizing the interaction around the condensates defined above
and introducing Nambu-Gorkov bispinors,
\beq
\Psi = \frac{1}{\sqrt{2}}\left(\begin{array}{c} q \\ C \bar{q}^T \end{array}
\right)\,,
\eeq
we obtain the mean-field Lagrangian
\beq
\mathcal{L}^{\mathrm{MF}} = \bar{\Psi} S^{-1} \Psi - \mathcal{V}
\eeq
where
\begin{alignat}{1}
  \mathcal{V} \;=\; & 2G\left(\phi_u^2+\phi_d^2+\phi_s^2\right) - 4
  K\phi_u\phi_d\phi_s
  \nonumber\\
  &+ \frac{1}{4H} \sum_{A,B = 2,5,7}
  \Big[\left({\Delta_{AB}^\scalar}\right)^2 +
  \left({\Delta_{AB}^\pscalar}\right)^2 \Big]
\label{eq:V}
\end{alignat}
is a field-independent potential density,
and $S^{-1}$ is the inverse dressed propagator, which in momentum space is 
given by
\begin{widetext}
  \beq
  S^{-1}(p) = \left(
    \begin{array}{cc}
      p\hspace{-1.4mm}/ + \hat\mu\gamma^0- \hat{M} & 
        \sum_{A,B} \left( \Delta_{AB}^\scalar \gamma_5 \tau_A \lambda_B 
        + \Delta_{AB}^\pscalar \tau_A \lambda_B \right) \\ 
      \sum_{A,B} \left(-\Delta_{AB}^\scalar \gamma_5 \tau_A \lambda_B +
        \Delta_{AB}^\pscalar \tau_A \lambda_B \right)& 
       p\hspace{-1.4mm}/  - \hat\mu\gamma^0 - \hat{M}
    \end{array}\right)\,.
  \label{eq:Sinv}
  \eeq  
\end{widetext}
It is a $72 \times 72$ matrix, corresponding to three color, three
flavor, four Dirac, and two Nambu-Gorkov degrees of freedom.
$\hat{M}$ is the diagonal matrix of the constituent quark masses
with the components
\beq
    M_a =  m_a - 4G\phi_a + 2K\phi_b\phi_c,
\label{eq:Ma}
\eeq
where $(a,b,c)$ is any permutation of $(u,d,s)$.

The quark part of the thermodynamic potential in 
mean-field  approximation is then given by
\beq
\Omega_q(T,\{\mu_i\}) = -T \sum_n \int
\frac{d^3p}{\left(2\pi\right)^3}\frac{1}{2} \Tr \ln \left(\frac{
  S^{-1}\left( i\omega_n,\vec{p}\right)}{T} \right) + \mathcal{V}.
\label{eq:OmegaMF}
\eeq
The $\Tr \ln$ is evaluated by determination of the 72 eigenvalues
of $S^{-1}$. 
Making use of a two-fold spin degeneracy, the problem can be reduced
to calculating the eigenvalues of a $36 \times 36$ matrix
\cite{Ruester:2005jc}. 
This matrix can be transformed further into block-diagonal form,
where the number and size of the blocks depend on the pairing pattern. 
For instance, in the CFL phase we get one $12\times12$ block and six 
$4\times4$ blocks, whereas in the CFL~$+$~Goldstone phases we 
have one $20\times20$ block and two $8\times8$ blocks.
The different blocks can be attributed to quasiparticle excitations with
different quantum numbers. For instance, in the \cflk phase the
$20\times20$ block corresponds to quasiparticles with $\tilde Q = 0$ 
whereas the two $8\times8$ blocks correspond to quasiparticles with
$\tilde Q = +1$ or $\tilde Q = -1$, respectively.

The resulting eigenvalues appear in the form 
$i\omega_n -  \epsilon_j(\vec p)$, where $\epsilon_j(\vec p)$ are the
energy-independent eigenvalues of the Matrix 
${\cal M} = i\omega_n - \gamma^0 S^{-1}$ and correspond to the 
dispersion relations of the quasiparticle modes. 
This allows for an analytic evaluation of the Matsubara sum, and we 
finally get
\beq
\Omega_q = - \int \frac{d^3p}{\left(2\pi\right)^3}
\sum_{j=1}^{18}\left(|\epsilon_j| +
  2\;T\;\ln\left(1+e^{-\frac{|\epsilon_j|}{T}}\right)\right)+\mathcal{V}.
\label{eq:Omega}
\eeq
Here we have used the fact that the eigenvalues come in pairs 
$(\epsilon_j,-\epsilon_j)$, so that we can restrict the sum, e.g., 
to the 18 positive eigenvalues, introducing an extra factor of 2.
The integral is divergent and will be regularized by a sharp 
three-momentum cutoff $\Lambda$.

In order to find the most stable solution, we have to minimize the
thermodynamic potential with respect to the various 
diquark and quark-antiquark condensates,
i.e., we have to solve the gap equations
\beq
\frac{\partial \Omega}{\partial \phi_f} =
\frac{\partial \Omega}{\partial \Delta_{AB}^\scalar} =
\frac{\partial \Omega}{\partial \Delta_{AB}^\pscalar} = 0\,,
\label{eq:gapeq}
\eeq
with $f\in\{u,d,s\}$ and $A,B \in \{2,5,7\}$.
In addition, since we are interested in quark matter under compact star 
conditions, our solution must fulfill the neutrality constraints
\beq
 n_Q = -\frac{\partial \Omega}{\partial \mu_Q} = 0,\quad
 n_3 = -\frac{\partial \Omega}{\partial \mu_3} = 0,\quad
 n_8 = -\frac{\partial \Omega}{\partial \mu_8} = 0.
\label{eq:neutr}
\eeq
\Eqs{eq:gapeq} and (\ref{eq:neutr}) form a 
set of coupled nonlinear equations which determines
the values of the condensates and the chemical potentials $\mu_Q$, $\mu_3$, 
and $\mu_8$ for any given quark chemical potential $\mu$ and temperature
$T$.
For the evaluation of the derivatives in these equations 
we follow \Ref{Warringa:2006dk}, 
making use of the diagonalized form of the propagator.
Finally, we checked for each solution that the off-diagonal color
densities vanish as well (cf.~\Ref{Buballa:2005bv} and the discussion in
Sec.~\ref{ssec:p2SC}). 

\subsection{Model parameters}

For the numerical calculations, we adopt the parameter values of 
Ref.~\cite{Rehberg:1995kh},
\begin{align}
  m_u = m_d &= 5.5 \MeV,\nonumber\\
  m_s &= 140.7 \MeV,\nonumber\\
  G &= 1.835 / \Lambda^2,\nonumber\\
  K &= 12.36 / \Lambda^5,\nonumber\\
  \Lambda &= 602.3 \MeV,
\end{align}
which have been fixed by fitting the vacuum values of the pseudoscalar 
meson masses and the pion decay constant.
The same parameter values have also been used in Refs.~\cite{Ruester:2005jc} 
and \cite{Warringa:2006dk}.
For the diquark coupling strength, which cannot be fixed in this way,
we follow Ref.~\cite{Ruester:2005jc} and take two different values,
namely $H= 3/4\;G$, which can be motivated by the Fierz transformation
of a one-gluon exchange, or $H=G$.

For the leptonic part, we consider massless electrons and muons with
mass 105.66~MeV. Here we differ slightly from 
\Refs{Ruester:2005jc,Warringa:2006dk} where a physical electron mass 
$m_e = 511$~keV was taken. Obviously this is negligible in most cases, 
but it could play a role at $T, |\mu_Q| \lesssim m_e$.
As we will see, this situation is just realized in the vicinity of the 
threshold of the gapless \cflk phase. We will comment on this at the end of 
Sec.~\ref{ssec:gCFLK}.

\section{Results}
\label{sec:results}

In this section we discuss the phase structure of the model. 
We begin with a brief overview and then analyze the most interesting
features in more detail.

\subsection{Phase diagram}
\label{ssec:phase_diagram}

\begin{figure*}[t!]
 \includegraphics[width=0.75\textwidth]{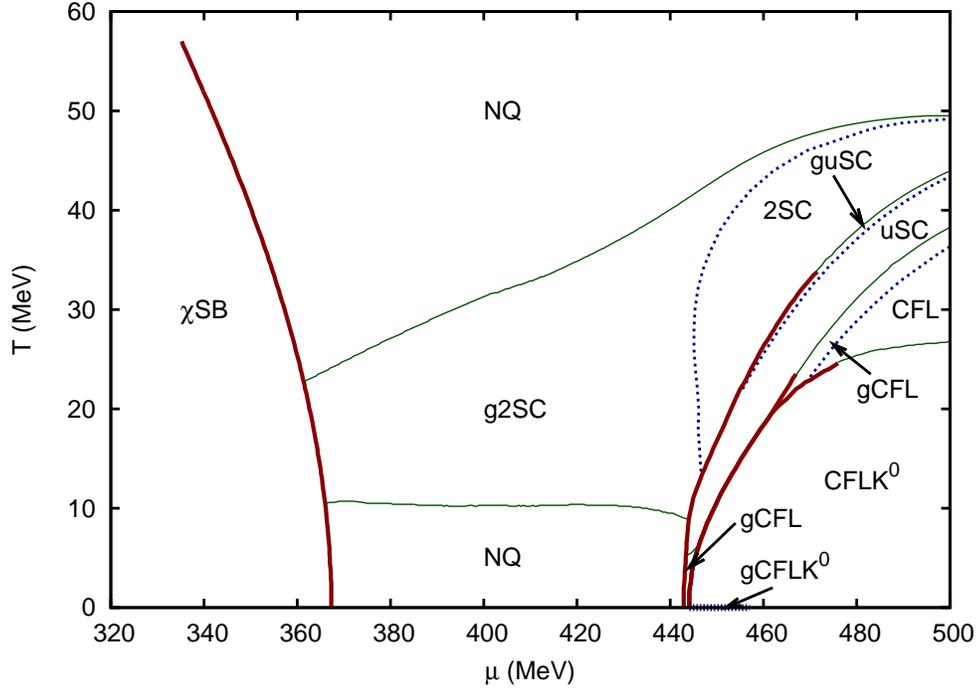}
 \caption{The phase diagram of charge and color neutral quark matter
   for ``intermediate diquark coupling'', $H = 3/4\;G$. Red thick
   solid lines denote first-order phase transitions, green thin solid
   lines second-order phase transitions, the blue dotted lines mark the
   (dis)appearance of the gap in the excitation spectrum.}
\label{phasediagram}
\end{figure*}
\begin{figure*}[t!]
  \includegraphics[width=0.75\textwidth]{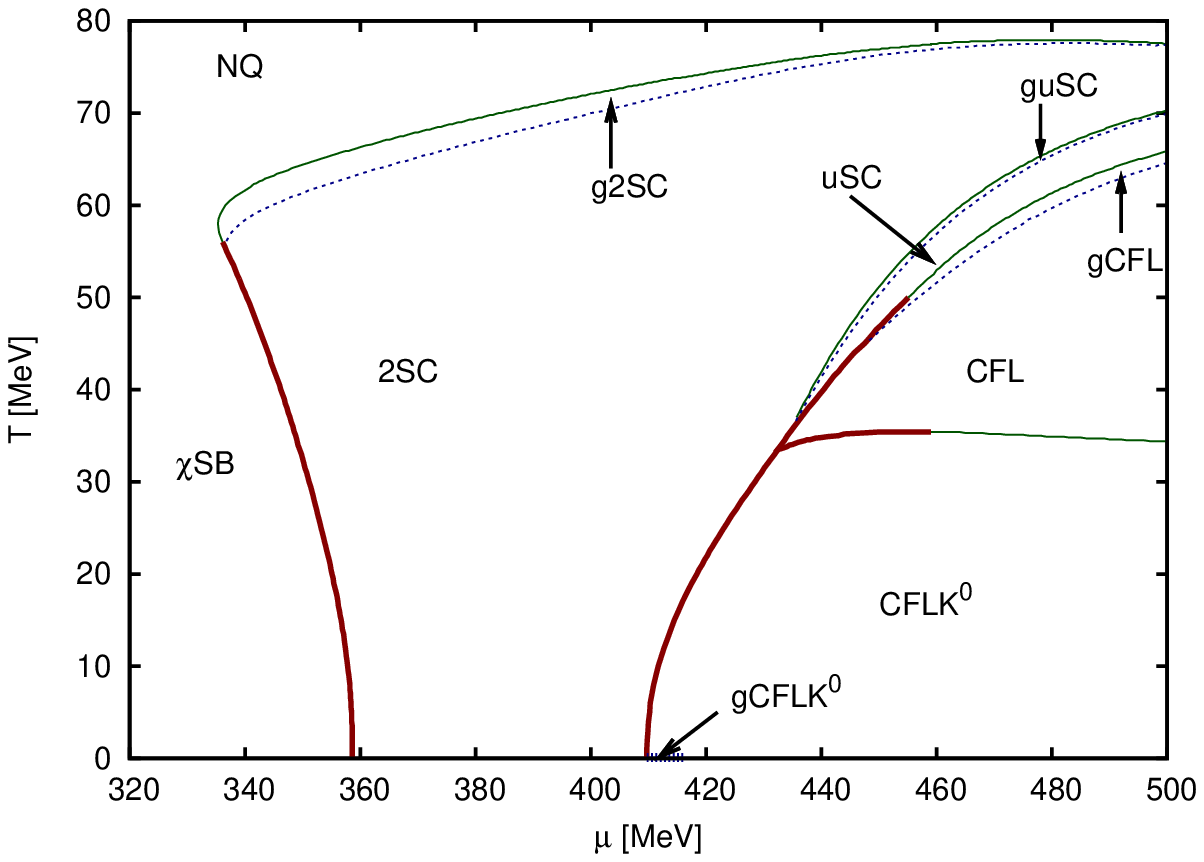}
\caption{The same as \Fig{phasediagram}, but 
  for ``strong diquark coupling'', $H = G$.}
\label{phasediagramHG}
\end{figure*}

As discussed in Sec.~\ref{ssec:Omega}, we obtain
the phase diagram in the $\mu - T$ - plane, by minimizing the
thermodynamic potential at every point of that plane 
with respect to the various condensates, taking into account the constraints 
of beta equilibrium and local electric and color neutrality. 
We thereby consider all phases which are listed in Tab.~\ref{phases}.
We recall that, while these phases are distinguished by their contents of 
nonvanishing diquark condensates, 
the chiral quark-antiquark condensates $\phi_f$ are selfconsistently
determined as well.
Strictly speaking, since chiral symmetry is explicitly broken in the model, 
this is not related to additional phases. However, following 
Ref.~\cite{Ruester:2005jc}, we label regions where $\phi_u$ and $\phi_d$
are large by `$\chi$SB' and regions where these condensates are small
by `NQ'. Both regimes are not necessarily separated by a phase boundary. 

The resulting phase diagrams are displayed in Figs.~\ref{phasediagram} 
and \ref{phasediagramHG}.
In Fig.~\ref{phasediagram} the diquark coupling constant was chosen to
be $H = 3/4\;G$. 
To a large extent our results agree qualitatively and quantitatively with 
those of Warringa, who has calculated the phase diagram within the same 
model for the same parameters \cite{Warringa:2006dk}:
Compared with \Ref{Ruester:2005jc}, where only scalar diquark condensates
have been taken into account, the main effect of considering pseudoscalar 
condensates is the existence of a \cflk phase, which replaces most of the 
(g)CFL phase at temperatures below $\approx 25\MeV$.
A second effect, which was reported
in \Ref{Warringa:2006dk}, is the emergence of a p2SC phase, i.e., a 
two-flavor color superconductor with an additional pseudoscalar condensate, 
replacing a part of the 2SC phase in the region around $\mu = 460$~MeV and 
$T = 30$~MeV. 
Although this result is confirmed by our calculations
(see \Fig{phasediagram_p2SC} below), we have not indicated
the p2SC phase in the phase diagram, because it is most likely a model
artifact. This will be discussed in Sec.~\ref{ssec:p2SC}.

While most of our results agree with 
\Ref{Warringa:2006dk}, there are basically two exceptions: 
First, we find that the \cflk phase becomes gapless at
extremely low temperatures and $\mu < 457.2$~MeV, 
whereas it was claimed in \Ref{Warringa:2006dk} that it is always gapped.
We discuss this in more details in Sec.~\ref{ssec:gCFLK}.
The second difference concerns the phase structure between the normal 
phase (NQ) and the \cflk phase at low temperatures. 
In contrast to the results of \Ref{Warringa:2006dk} we do not find a uSC
phase down to zero temperature. Instead we find a situation more like in
\Ref{Ruester:2005jc}, where at temperatures lower than $\approx 7\MeV$ a 
gCFL phase is realized.
This will be discussed in Sec.~\ref{ssec:gCFL}.

In Fig.~\ref{phasediagramHG} we show the phase diagram for the diquark 
coupling constant $H=G$.
This case was not studied in \Ref{Warringa:2006dk}.
However, it was investigated in \Ref{Ruester:2005jc}, considering scalar
diquark condensates only. By comparing that phase diagram with 
Fig.~\ref{phasediagramHG}, we see that the effects of the pseudoscalar 
condensates are qualitatively similar to the previous case:
A major part of the CFL phase is replaced by the \cflk phase, which 
now persists up to temperatures of about $35\MeV$.
At very low temperatures and $\mu < 415.6$~MeV the \cflk phase is gapless. 
Numerically, we also find again a p2SC phase, which replaces the 2SC phase 
in some region. However, as mentioned before,
we believe that this is a model artifact (see Sec.~\ref{ssec:p2SC})
and do not indicate this phase in the phase diagram.

\subsection{The gapless \cflk phase}
\label{ssec:gCFLK}

\begin{figure}[!]
  \includegraphics[width=0.47\textwidth]{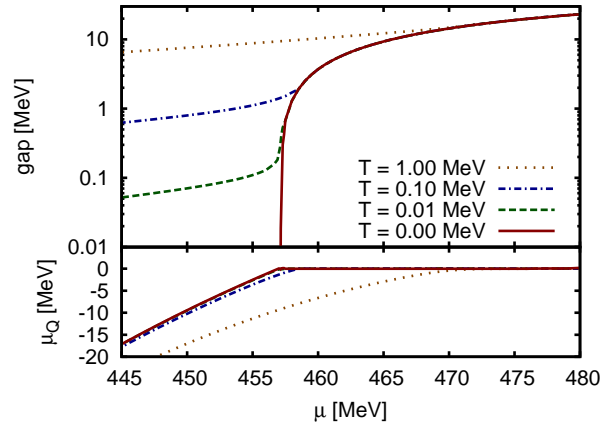}
  \caption{The gap in the excitation spectrum and the electric charge
    chemical potential $\mu_Q$ in the (g)\cflk phase 
    as functions of the chemical potential for $H = 3/4\;G$ 
    and various temperatures.}
  \label{gapmuQ}
\end{figure}

Both, in \Fig{phasediagram} and \Fig{phasediagramHG}, there is
a gapless \cflk phase at very low temperatures below a certain value 
of $\mu$.
For the case $H = 3/4\;G$ this is illustrated in \Fig{gapmuQ}.  
In the upper part, the gap in the quasiparticle excitation 
spectrum is displayed as a function of $\mu$.
In general, the gaps are nonzero at higher chemical potentials
and decrease with decreasing chemical potential. 
At zero temperature (red solid line) it eventually goes to zero at
$\mu = 457.2$~MeV, so that there is indeed a gapless phase below 
that threshold.

It turns out, however, that this behavior is extremely sensitive 
to the temperature.
At $T=1$~MeV (yellow dotted line), which is often a very good 
approximation to zero temperature, the gap stays of the order of several
MeV and there is no indication for a nearby gapless phase. 
At lower temperatures, the gap first follows the $T=0$ behavior
but then levels off and stays finite.
Even at $T=0.01$~MeV (green dashed line), the gapless phase is not 
reached. This seems to suggest, that the gapless \cflk phase only exists 
at zero temperature.\footnote{Note that this has
nothing to do with the fact that only at $T=0$ the transition from a 
gapped to a gapless phase corresponds to a real phase transition with
a well defined order parameter. Here we identify the gapless regimes
by inspecting the gaps in the excitation spectrum, which can be done
at any temperature. As one can see in \Figs{phasediagram} and 
\ref{phasediagramHG}, other gapless phases defined in this way do exist 
at $T>0$, see also 
\Refs{Ruester:2005jc,Blaschke:2005uj, Abuki:2005ms,Warringa:2006dk}.}

\begin{figure}
  \includegraphics[width=0.47\textwidth]{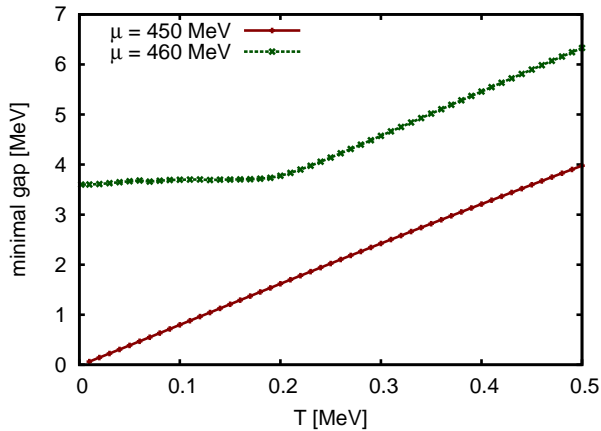}
  \caption{The gap in the excitations spectrum as function of the 
    temperature for $H = 3/4\;G$ and two different values of $\mu$.}
  \label{gapT}
\end{figure}

This picture is corroborated by \Fig{gapT}, where the excitation gaps are 
displayed as functions of temperature at fixed chemical potential. 
For $\mu = 450$~MeV the numerical values
(which are explicitly indicated by the points)
show an almost perfect linear behavior,
so that, within numerical accuracy, the 
\cflk phase becomes gapless exactly at $T=0$.
For comparison, we also show the results for $\mu = 460$~MeV. 
In this case the gap levels off and stays finite at $T=0$.

In the lower part of \Fig{gapmuQ} the electric charge chemical 
potential $\mu_Q$ is plotted as a function of $\mu$ for the same 
temperatures as the excitations gaps shown in the upper part. 
At zero temperature, as explained in Sect.~\ref{ssec:phases},
$\mu_Q$ is an order parameter, which vanishes in the
gapped phase and is nonzero in the gapless phase.
This is no longer true at $T>0$ where $\mu_Q$ is always nonzero.
Nevertheless, one might expect that at low enough temperatures $\mu_Q$
is very small in the gapped regime, so that larger deviations from zero
signal a gapless phase. It turns out, however, that this is not the case.
For $T=1$~MeV, $0.1$~MeV, or $0.01$~MeV, $\mu_Q$ exhibits a threshold-like 
behavior, although we do not enter a gapless regime.
When we compare these ``thresholds'' with the upper part 
of \Fig{gapmuQ}, we see that they more or less coincide
with the positions where the excitation gaps start deviating from
the $T=0$ gaps.  In fact, as we will discuss,
the mechanism which keeps the excitation gaps finite at $T>0$ is
closely related to the rise of $-\mu_Q$.

To understand this, we take a more detailed look at the low-energy
quasiparticle excitation spectra and thereby compare the 
g\cflk phase with the more familiar gCFL phase.
A similar comparison has been performed earlier
in \Ref{Forbes:2004ww}, but restricted to zero temperature and
to the gapped or the threshold to the gapless regime. 
Here we go deeper into the gapless regime and also discuss temperature
effects.

\begin{figure*}
  \includegraphics[width=0.47\textwidth]{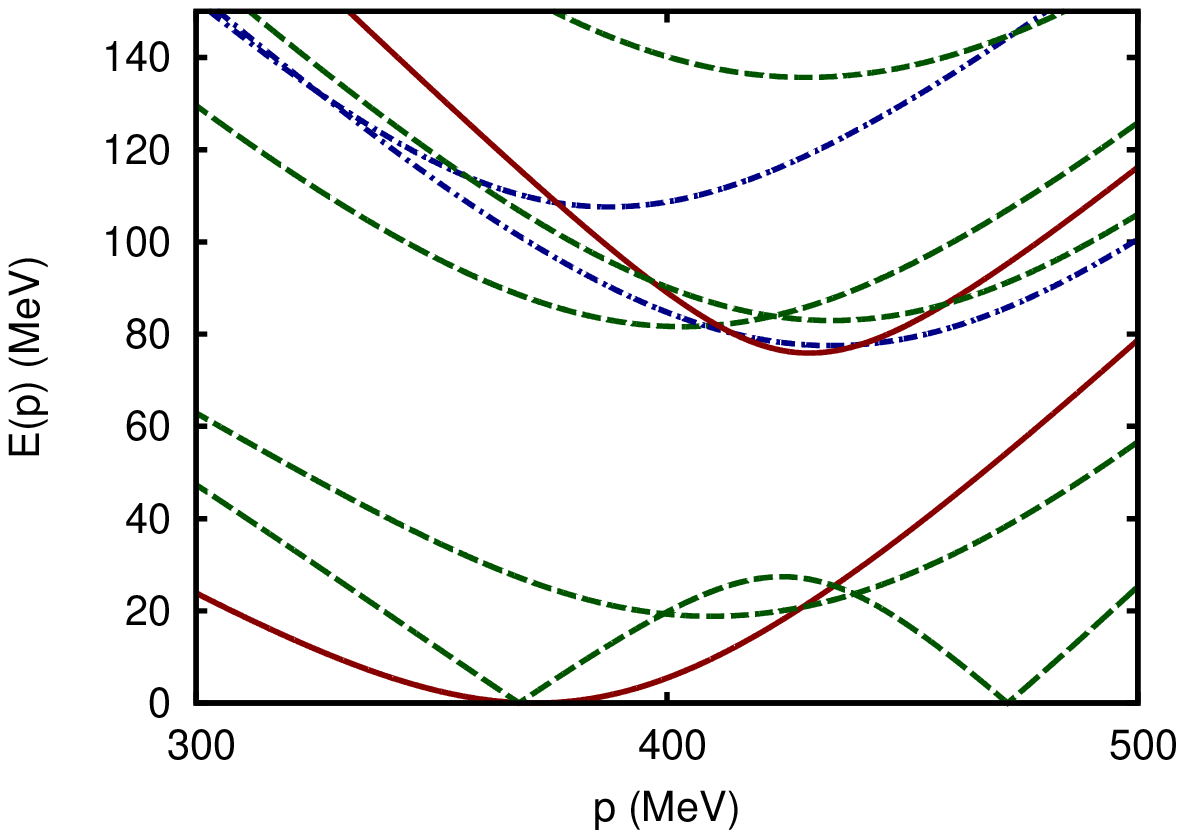}
  \includegraphics[width=0.47\textwidth]{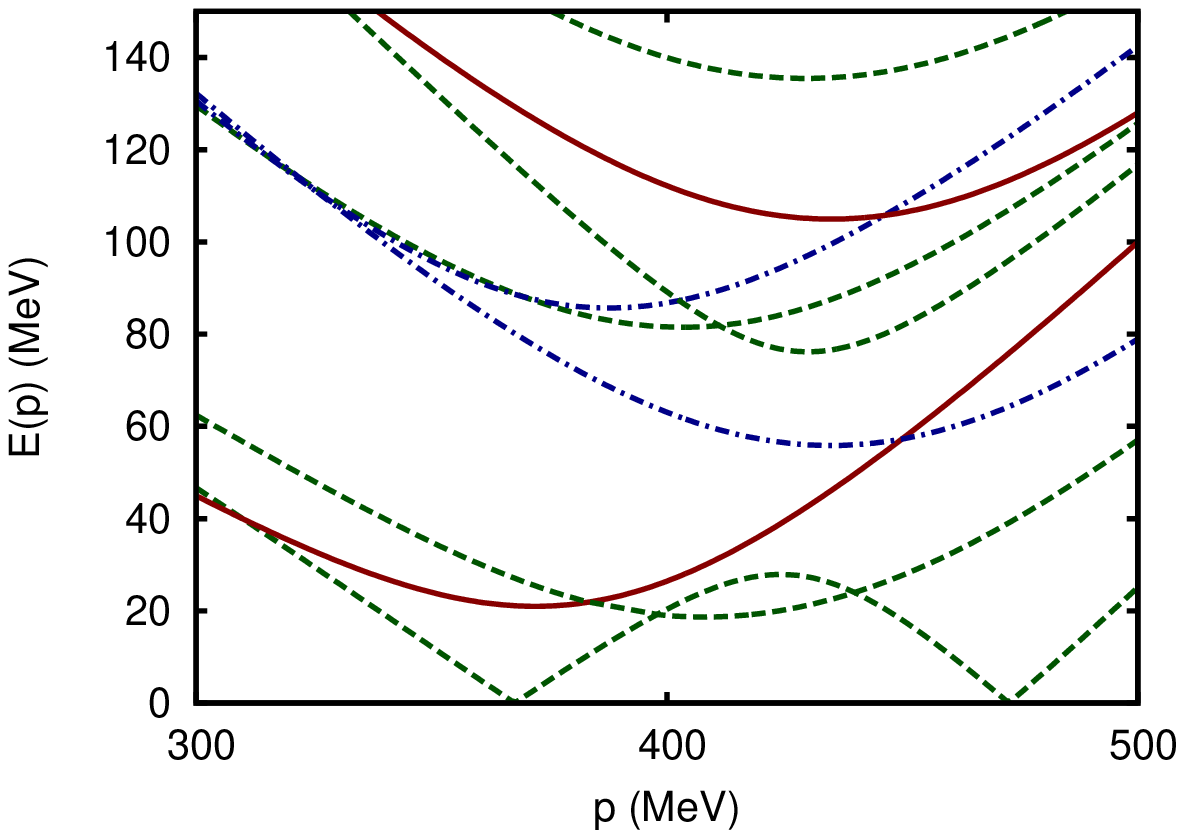}
  \includegraphics[width=0.47\textwidth]{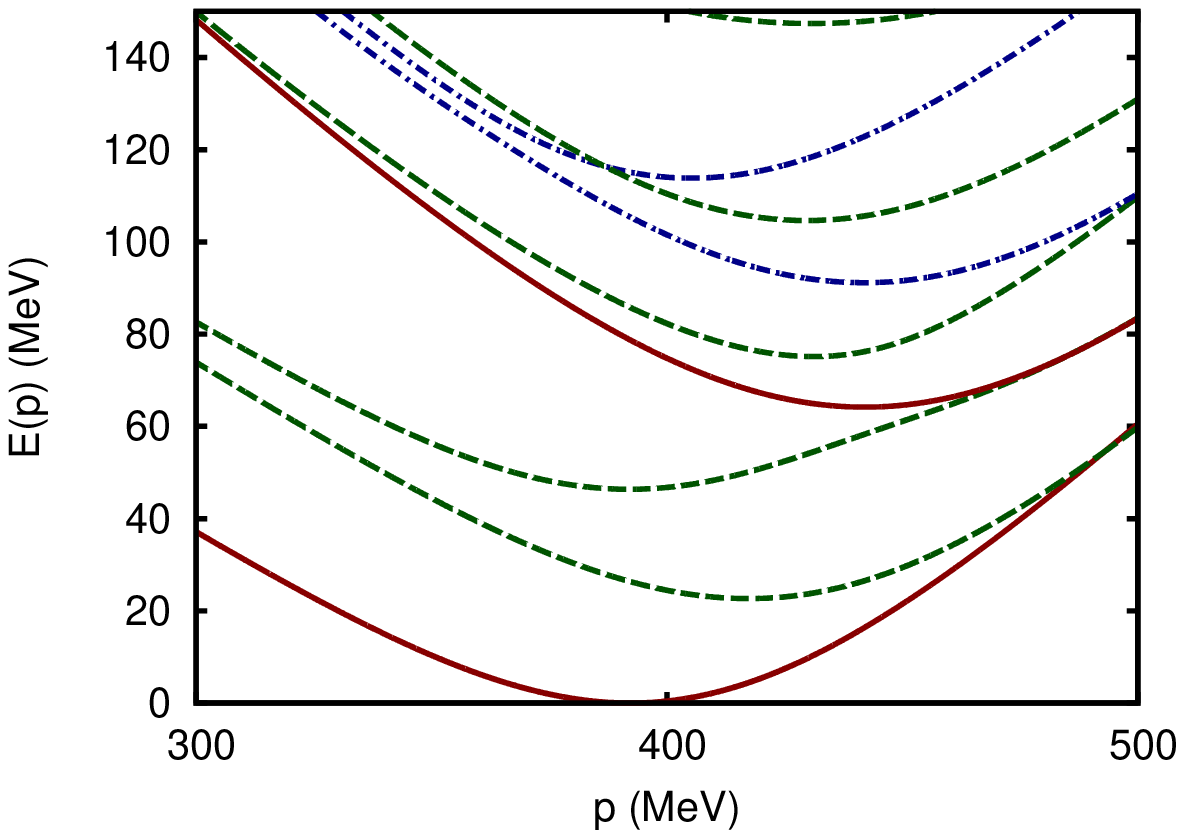}
  \includegraphics[width=0.47\textwidth]{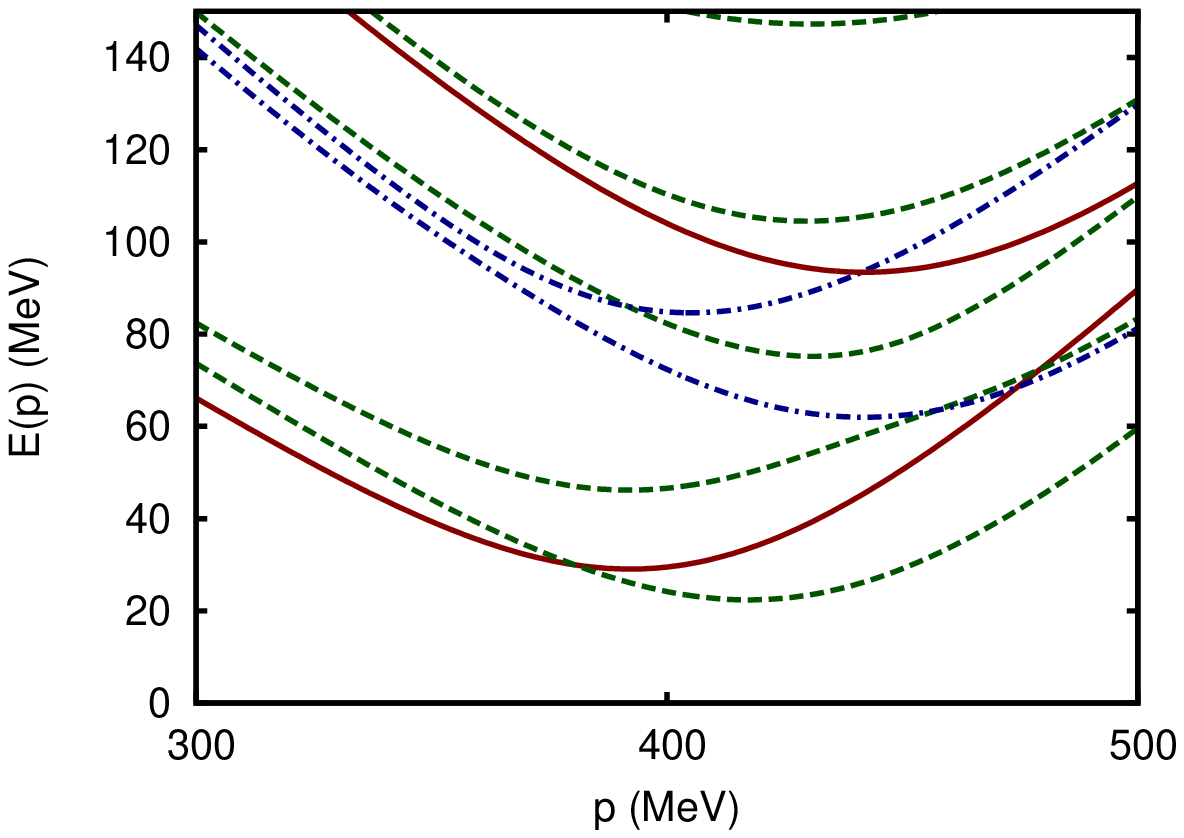}
  \caption{Quasiparticle dispersion relations in the gCFL phase (top)
    and in the (g)\cflk phase (bottom) at $\mu = 450$~MeV and
    $T = 0$ (left) or $T = 5$~MeV (right); $H = 3/4\;G$. 
    The different line types correspond to modes with different 
    $\tilde Q$ charges: 
    $\tilde Q = +1$ (red solid lines),
    $\tilde Q = 0$ (green dashed lines),
    $\tilde Q = -1$ (blue dash-dotted lines).
  }
  \label{disp}
\end{figure*}

In \Fig{disp} the quasiparticle spectra at $\mu = 450$~MeV are shown
for $T=0$ (left) and $T=5$~MeV (right).  The two lower panels
correspond to the (g)\cflk phase, while in the two upper panels the
analogous gCFL solutions are shown, which are obtained if the
pseudoscalar diquark condensates are not taken into account.  Since
$\tilde Q$ is a conserved quantum number in both cases, the
quasiparticle excitations carry a definite $\tilde Q$ charge. These
$\tilde Q$ charges, which can be $0$, $+1$, or $-1$, are indicated in
\Fig{disp} as well.

Let us begin with the discussion of the gCFL solutions. 
At $T=0$ (upper left panel) there are two gapless modes. 
One of them vanishes at two different momenta,
separated by a ``blocking region'' of about 100~MeV.
This mode has $\tilde Q = 0$ and can be identified with a $bd$-particle 
$gs$-hole mode. In addition, there is another gapless mode
(which has $\tilde Q = +1$ and is a $bu$-particle $rs$-hole mode),
which seems to vanish just at its minimum. 
On a much smaller scale one finds that this is not exactly true,
but there is a finite blocking region as well. It is, however,
very small and therefore not visible in the figure.
(For the analogous mode in the g\cflk phase, this is discussed in 
\Fig{blocking}.)

This behavior is well known and has been explained in
\Ref{Alford:2004hz}.  It was shown that at $T=0$ both modes, the
$bd-gs$ mode and the $bu-rs$ mode, become gapless at the same critical
chemical potential but behave rather differently when $\mu$ is further
decreased: In the $bu-rs$ mode, the pair breaking leads to unpaired
$bu$ quarks in the blocking region, which carry positive $\tilde Q$
charge. Color neutral quark matter is therefore positively charged and
must be neutralized by electrons. This requires $\mu_Q < 0$, which, in
turn, reduces the stress in the $bu-rs$ pairs and keeps the blocking
region very small.  For the $bd-gs$ mode, on the other hand, there is
no such mechanism because both quarks have $\tilde Q = 0$.  As a
consequence, the blocking region becomes much wider.

When we increase the temperature to $5$~MeV, as shown in the upper
right panel of \Fig{disp}, the $bd-gs$ mode remains gapless, whereas
the $bu-rs$ mode receives a considerable gap.  A detailed numerical
analysis reveals that this mode behaves in a similar way as the lowest
\cflk mode in \Fig{gapT}.  In particular, it becomes already gapped if
the temperature is increased only slightly above zero.  This is
naturally explained by Fermi smearing effects.  In fact, it is a
well-known phenomenon that modes which are gapless at $T=0$ can become
gapped above a certain temperature~\cite{Huang:2003xd}.  In the
present case, where the blocking region at $T=0$ is very small, it is
plausible that the gapless regime is restricted to very small
temperatures.  On the other hand, since the size of the blocking
region at $T=0$ is not exactly equal to zero, the ``critical''
temperature should be nonzero as well. It is, however, too small to be
determined numerically.

\begin{figure}
  \includegraphics[width=0.47\textwidth]{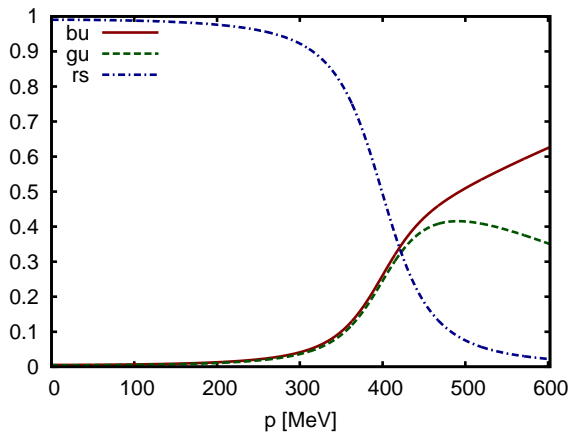}
  \caption{Composition of the gapless $\tilde Q = 1$ mode
    in the \cflk phase for $H = 3/4\;G$ at $\mu = 450$~MeV and $T = 0$ 
    (cf.~\Fig{disp}, lower left panel).
    Shown are the squared coefficients of the color-flavor
    components of the eigenfunctions as functions of the three-momentum.}
  \label{composition}
\end{figure}

\begin{figure}
  \includegraphics[width=0.47\textwidth]{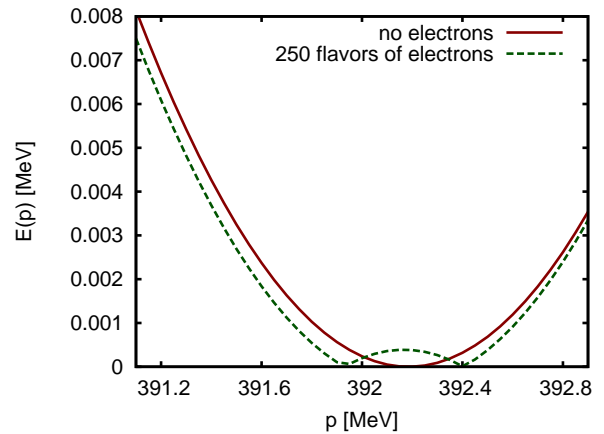}
  \caption{Dispersion relation of the gapless $\tilde Q = 1$ mode 
    in the \cflk phase at $\mu = 450$~MeV and $T = 0$ for $H = 3/4\;G$. 
    Shown are the
    neutral solutions in a world with no electrons (red solid line)
    and in a world with 250 independent electron species (green
    dashed line). The solution in the real world with one electron
    species would be very close to the solid line, but qualitatively
    look like the dashed line, with a tiny blocking region.
  }
  \label{blocking}
\end{figure}

We now compare these results with the spectra in the (g)\cflk phase,
which are displayed in the lower part of \Fig{disp}.  At $T=0$ (left),
there is again a marginally gapless $\tilde Q = +1$ mode with a tiny
blocking region around its minimum (cf.~\Fig{blocking} below).  The
essential difference to the gCFL phase is, however, that the g\cflk
phase contains no second gapless mode, in which $\tilde Q = 0$ quarks
are paired.  The size of the gapless regime is therefore determined by
the lowest $\tilde Q = +1$ mode.  The latter can be seen as a ``close
relative'' of the $bu-rs$ in the gCFL phase.  In the (g)\cflk phase a
complication arises due to the fact that not only two, but four quark
species mix in the $\tilde Q=\pm 1$ quasiparticle modes
(cf.~\Fig{pairing}).  The result of the numerical diagonalization is
shown in \Fig{composition} where the composition of the gapless mode
is displayed as a function of the three-momentum. In agreement with
\Ref{Forbes:2004ww}, we find that the gapless mode is mainly a mixture
of $bu$ and $gu$ particles with $rs$ holes.  The blocking of the $rs$
quarks in the gapless regime then leads again to a surplus of positive
$\tilde Q$ charge, which forces the blocking region to stay very
small.  As a consequence, this mode and, hence the \cflk phase is
gapless only at very low temperatures.

Our results at $T=0$ are in qualitative agreement with the effective
field theory prediction that the first gapless mode in the \cflk phase
is electrically charged~\cite{Kryjevski:2004jw, Kryjevski:2004kt,
  Gerhold:2006np}.  According to these works, there could be a second
phase transition at lower chemical potential, where a neutral mode
becomes gapless as well. In fact, we found such solutions in some
region of the phase diagram, but they were never favored. We will come
back to this point in Sec.~\ref{ssec:gCFL}.

Concerning the electrically charged gapless modes, we found some
quantitative differences to the effective theory results.  In
\Refs{Kryjevski:2004jw,Kryjevski:2004kt} it was predicted that the
charged mode becomes gapless at $M_s^2/(2\mu) \approx 4/3~\Delta$,
where $\Delta$ is the CFL gap in the chiral limit. Comparing this with
the corresponding threshold in the CFL phase, $M_s^2/(2\mu) \approx
\Delta$, and assuming that the strange quark mass is the same in both
cases, this means that the critical chemical potential for entering
the gapless regime is 25\% lower in the kaon condensed phase.  In
\Ref{Forbes:2004ww} it was found that the effect is somewhat weaker in
an NJL model with $\Delta=25$~MeV.  It turns out that this tendency
continues with increasing coupling strength: For $H=3/4\;G$ we find
that both transitions take place at approximately the same chemical
potential, while for $H=G$ the critical chemical potential for
entering the gapless regime is even shifted upwards by almost 10~MeV
in the kaon condensed phase.  Partially, this can be attributed to the
fact that the selfconsistently calculated strange quark mass in the
\cflk phase is a few percent larger than in the CFL phase, so that the
rotation into the kaon condensed phase reduces the stress not as much
as for a constant mass.  But even if we take the mass ratio out, there
remains a discrepancy.  This indicates that there are higher-order
effects which have been neglected in the effective theory approach,
but become important at stronger couplings.

We would like to note that the numerical solution of the coupled gap
equations (\ref{eq:gapeq}) together with the neutrality conditions
(\ref{eq:neutr}) turned out to be extremely difficult to obtain in the
g\cflk regime. As obvious from the behavior of $\mu_Q$ at $T=0$, an
extrapolation of the solutions from the gapped regime to the gapless
regime does not work.  Formally, this is related to the fact that,
whereas for the gapped solutions the thermodynamic potential at fixed
chemical potentials is a minimum with respect to the various
condensates, for the gapless solutions it is a maximum with respect to
$\Delta_{55}^\scalar$, $\Delta_{52}^\pscalar$ and $\phi_s$.  At the
unpairing line, these minima and maxima merge, and the solutions
correspond to inflection points in these parameters.\footnote{This was
  also found in Ref.\cite{Alford:2004hz} for $\Delta_{55}^\scalar$ in
  the gCFL phase.}  As a consequence, a small variation of the
chemical potential can easily lead to a situation, where the solutions
of the gap equations disappear completely. Obviously, that makes an
iterative solution of the problem in this regime practically
impossible.

Since the gapless solutions are very close to the unpairing line,
these difficulties are hard to avoid in realistic cases.  We have
therefore followed the idea of \Ref{Alford:2004hz} and introduced
additional independent ``flavors'' of electrons.  Then, in order to
neutralize the extra negative charge of these electrons, a larger
blocking region is needed, so that the gapless solutions move away
from the unpairing line and the solutions are easier to find.  The
resulting dispersion relation of the gapless mode with 250 electron
flavors at $T=0$ and $\mu = 450$~MeV is indicated by the green dashed
line in Fig.~\ref{blocking}. We see that this mode has a blocking
region of about 0.5~MeV. For comparison we also show the solution for
a world without electrons (red solid line). In this case, there are no
negative particles which could neutralize the positive charge of
unpaired up quarks.  Therefore the size of the blocking region must
vanish, i.e., the solution is exactly located on the unpairing
line. In this case it can be found coming from the gapped regime.

On the scale of Fig.~\ref{disp}, the two cases shown in
Fig.~\ref{blocking} would be indistinguishable and, hence,
indistinguishable from the real case with one electron flavor.  We
also found that the solutions for the gap parameters, masses, and
chemical potentials obtained with 250 or no electron flavors differ
only little.  The calculation with 250 electron flavors in
Fig.~\ref{blocking} mainly serves as an example to prove that gapless
\cflk solutions with a finite blocking region indeed exist.
Otherwise, of course, the solutions without electrons should be closer
to the realistic case.  For most practical purposes, we can therefore
well approximate the latter by a calculation without electrons.  In
fact, this is what we have done in Figs.~6 and~7, and in Fig.~4 for
$\mu_Q$ at $T=0$.  However, we repeat that the solutions for 250
electron flavors and, hence, for one electron flavor would look the
same.

Finally, we want to comment on the role of a finite electron mass $m_e
= 511$~keV, which we have neglected so far.  Since $T \ll m_e$ in the
g\cflk phase, this is only justified in the regime where $|\mu_Q| \gg
m_e$.  As we can see in Fig.~\ref{gapmuQ}, this condition is fulfilled
in the major part of the g\cflk phase.  In particular, the existence
of the g\cflk phase is not precluded by the electron mass.  Of course,
$m_e$ does have a quantitative effect near the threshold to the gapped
regime and also on the threshold itself.  We recall that, because of
$\tilde Q$ neutrality, color neutral quark matter is automatically
electrically neutral in the gapped \cflk phase at $T=0$.  Therefore no
electrons are admitted in the gapped regime, because they would spoil
the overall neutrality.  However, for nonzero $m_e$ this does no
longer force $\mu_Q$ to vanish, but any $|\mu_Q| < m_e$ is allowed.
As a consequence, the system can stay in the gapped phase until the
unpairing line, which in the relevant regime is given by the red solid
line in the lower part of Fig.~\ref{gapmuQ}, crosses the line $\mu_Q =
-m_e$.  This shifts the threshold to the gapless phase downwards in
$\mu$ by about 0.4~MeV.

\subsection{The gCFL window between normal and \cflk phase}
\label{ssec:gCFL} 

We have seen in \Figs{phasediagram} and \ref{phasediagramHG} that
large parts of the CFL or gCFL phase at not too high temperatures are
replaced by the \cflk phase when pseudoscalar condensates are taken
into account~\cite{Warringa:2006dk}.  As argued in the Introduction,
the $K^0$ condensate reduces the number of strange quarks, which is
favorable at ``moderate'' densities, where the strange quark mass
cannot be neglected~\cite{Schafer:2000ew,Bedaque:2001je}.  Eventually,
at very high chemical potentials, the CFL phase should be the most
favored phase, but this happens far beyond the range of applicability
of our model, which is limited by the cutoff.  From this perspective
it appears somewhat surprising that for $H = 3/4~G$ the gCFL phase
survives in a small window at the lower-$\mu$ end of the \cflk phase
and low temperatures, see \Fig{phasediagram}.  This result is also at
variance with \Ref{Warringa:2006dk} where a uSC phase was found in
this regime.


\begin{figure}
  \includegraphics[width=0.47\textwidth]{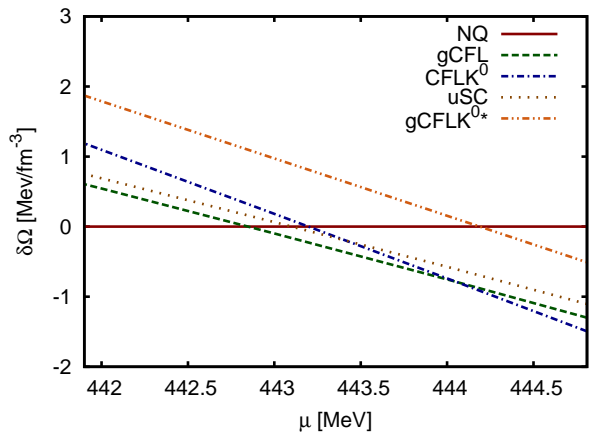}
  \caption{The free energy of different solutions relative to 
    the normal phase, 
    $\delta\Omega(T,\mu) = \Omega(T,\mu) - \Omega_{NQ}(T,\mu)$,
    for $H = 3/4\;G$ and $T=0.5$~MeV as a function of $\mu$.} 
  \label{dOmega}
\end{figure}


Our result is illustrated by \Fig{dOmega}, where the free energy of
different solutions relative to the normal phase, $\delta\Omega =
\Omega - \Omega_{NQ}$, at $T = 0.5$~MeV is displayed against $\mu$ in
the interesting regime. While the normal phase and the \cflk phase
have the lowest free energy at low or high chemical potential,
respectively, we see that there is indeed a small intermediate regime
where the gCFL phase is favored. The uSC phase, on the other hand, is
never favored.

We have mentioned earlier that for low enough chemical potentials,
effective field theories predict the existence of \cflk solutions with
neutral gapless modes. In the following we will call these solutions
`g\cflks solutions' for a better distinction from the ``ordinary''
\cflk solutions with charged gapless modes, as discussed in
Sec.~\ref{ssec:gCFLK}. In fact, at $T=0$, the g\cflks solutions
contain both, neutral and charged gapless modes. However, as we have
seen for the gCFL phase, the neutral gapless mode should be far more
stable at finite temperature.

In Ref.~\cite{Gerhold:2006np} it was found that at $T=0$ the g\cflks
phase is reached from the ordinary g\cflk phase in a first-order phase
transition, where the condensate $\Delta_{77}^\scalar$ drops
substantially. This is not very different from our \cflk $\rightarrow$
gCFL phase transition, which is also first order and accompanied by a
strong drop of $\Delta_{77}^\scalar$. In fact, the authors of
Ref.~\cite{Gerhold:2006np} found that near the phase transition the
free energy of the g\cflks phase is very close to the free energy of
the gCFL phase, so that the result that the former is favored is not
very robust.

We have therefore searched for g\cflks solutions in this regime. To
that end, we have rotated the gCFL solution into the $K^0$ direction
and used the resulting gap parameters as starting values for the
iterative solution of the coupled gap equations and neutrality
conditions. In this way, we indeed found g\cflks solutions, but they
turned out not to be favored. For $T=0.5$~MeV, this is shown in
\Fig{dOmega} as well.


\begin{figure}
  \includegraphics[width=0.47\textwidth]{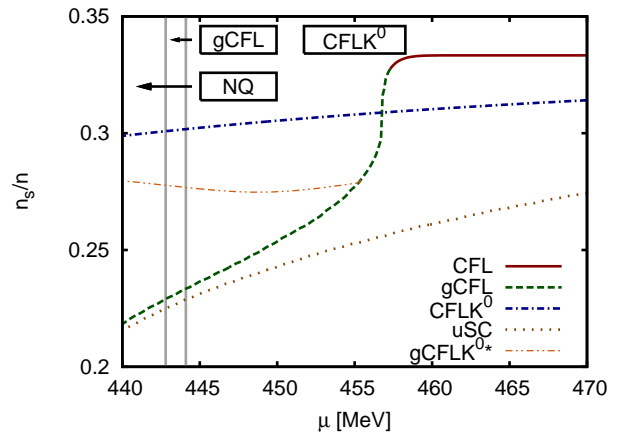}
  \caption{The fraction $n_s/n$ of strange quarks in different color 
    superconducting solutions 
    for $H = 3/4\;G$ and $T=0.5$~MeV as a function of $\mu$.
    In the boxes it is indicated which phase is realized in the phase 
    diagram (\Fig{phasediagram}).} 
  \label{ns_n}
\end{figure}


In order to shed some light on these results, we compare the fraction
of strange quarks in the various solutions. This is shown in
\Fig{ns_n}.  In the fully gapped CFL phase at zero temperature $n_s/n
= 1/3$, due to the absence of electrons~\cite{Rajagopal:2000ff}.  At
$T = 0.5$~MeV, this does no longer hold exactly. Nevertheless, as one
can see in the figure, $n_s/n$ is still very close to one third (solid
line).  Below $\mu = 457.1$~MeV the CFL solution becomes gapless
(dashed line).  Near this point the strange quark fraction drops very
steeply with decreasing chemical potential and then approaches
smoothly the uSC value (dotted).

The fraction of strange quarks in the \cflk solution is indicated by
the dash-dotted line. As expected, the $K^0$ condensation leads to a
reduction of the strangeness content as compared to the fully gapped
CFL solution.  However, when going to lower chemical potentials,
$n_s/n$ decreases only mildly and is therefore considerably larger
than the value in the gCFL solution. From this point of view, it is
less surprising that the gCFL solution eventually gets favored.  Of
course, the strangeness content is only one aspect.  The reduction of
strangeness in the gCFL phase is related to the breaking of pairs and,
thus, comes together with a reduction of the pairing energy.  As a
result, the \cflk phase persists to rather low chemical potentials and
the actual phase structure strongly depends on details of the
interaction.

For completeness we also show the strangeness fraction in the g\cflks
solution (dash-double dotted line). This solution only exists below
$\mu = 455.3$~MeV, where it diverges from the gCFL solution by
developing pseudoscalar condensates.  At lower $\mu$ it approaches the
gapped \cflk solution, and we have strong indications that these
solutions eventually join.  However, the numerical analysis becomes
rather difficult in this regime and, since both solutions are
disfavored anyway, we did not make much effort to work this out more
carefully.

\subsection{The p2SC phase}
\label{ssec:p2SC}


\begin{figure}
  \includegraphics[width=0.48\textwidth]{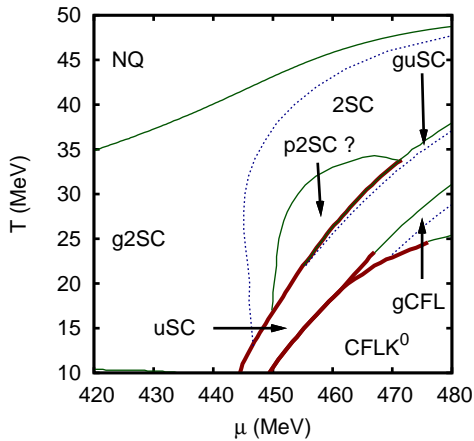}
  \caption{Detail of the phase diagram for $H = 3/4\;G$. In 
    contrast to \Fig{phasediagram} we have also indicated the region
    where we numerically find the p2SC phase to be favored.} 
  \label{phasediagram_p2SC}
\end{figure}


As mentioned in Sec.~\ref{ssec:phase_diagram}, we find numerically
that the 2SC phase gets partially replaced by a p2SC phase when
pseudoscalar condensates are taken into account, see
\Fig{phasediagram_p2SC}.  This was first observed in
\Ref{Warringa:2006dk}.  It was found there that in those regions where
the p2SC phase is favored, the free energy gain compared with the 2SC
solution is very small, less than 1~keV/fm$^3$.  It was also noted
that the p2SC solution is connected to the corresponding 2SC solution
via an axial color transformation, \beq q \rightarrow \exp(i \theta
\frac{\lambda_7}{2}\gamma_5)q\,,
\label{eq:axcol}
\eeq
which partially rotates the condensate $\Delta_{22}^{\scalar}$ into
$\Delta_{25}^{\pscalar}$.
In this case the condensates in the two phases should be related as
\beq 
\left(\Delta_{22}^{\scalar}\right)_{\text{p2SC}}^2 +
\left(\Delta_{25}^{\pscalar}\right)_{\text{p2SC}}^2
=
\left(\Delta_{22}^{\scalar}\right)_{\text{2SC}}^2\,. 
\label{eq:condaxrot}
\eeq
Indeed, the numerical solutions were found to be in very good agreement with 
this relation. We confirm all these results. 

Nevertheless, the physical significance of this phase is not clear.
Unlike the \cflk phase, the p2SC phase is not obviously related to the
condensation of a Goldstone mode. In fact, the axial color rotation,
\Eq{eq:axcol}, is neither a symmetry of QCD nor of our NJL-model
Lagrangian, \Eq{L}, not even in the chiral limit.  On the other hand,
for $M_u = M_d = 0$ and vanishing color chemical potentials, the
mean-field thermodynamic potential is invariant under \Eq{eq:axcol} if
the transformation is restricted to the non-strange sector and only
the diquark condensates $\Delta_{22}^{\scalar}$ and
$\Delta_{25}^{\pscalar}$ are present (see App.~\ref{app:p2SCsym}).  As
a consequence, there is a spurious pseudoscalar Goldstone mode in the
2SC phase at mean field, which is related to the spontaneous breaking
of this accidental symmetry by $\Delta_{22}^{\scalar}$.  This suggests
that the p2SC phase is an artifact of the mean-field approximation.

Yet, the accidental symmetry does not explain, why the partial rotation 
of $\Delta_{22}^{\scalar}$ into $\Delta_{25}^{\pscalar}$ becomes favored
in a certain regime, when we turn to the physical case with nonvanishing
up and down quark masses and color chemical potentials. 
In this context, it is quite instructive to make a comparison with
the case of an ordinary (vector) SU(3) color rotation,  
\beq
q \rightarrow \exp(i \theta_a \frac{\lambda_a}{2})q\,.
\label{eq:veccol}
\eeq It has been observed some time ago that, seemingly, this symmetry
can be exploited to neutralize the 2SC phase without introducing color
chemical potentials~\cite{Blaschke:2005km}.  Thereby
$\Delta_{22}^{\scalar}$ is rotated into a different direction in color
space, so that equally many red, green, and blue quarks are paired in
the condensate.  As a consequence, the color charge densities $n_3$
and $n_8$ vanish.  Moreover, since no color chemical potentials are
needed, the resulting state has a lower free energy than the standard
neutral 2SC phase with nonvanishing $\mu_8$.  However, as discussed in
\Ref{Buballa:2005bv}, this result is not correct if the NJL model is
taken as a model for QCD: A colored state can of course not be
neutralized by a color rotation.  The color charge is just rotated
into a different direction, and instead of $n_8$ one finds
nonvanishing off-diagonal color charge densities $n_a = \av{q^\dagger
  \lambda_a q}$ with $a = 1,4,6$ \cite{Buballa:2005bv}.  In QCD, these
color charge densities couple to the corresponding gluon fields
$A^0_a$, which then neutralize the system for all color components.
In NJL-type models, where no gluons are present, this must be mimicked
by introducing the corresponding (off-diagonal) color chemical
potentials.  Using these color chemical potentials to remove all
(diagonal and off-diagonal) color charge densities, the resulting
state is completely equivalent to the standard 2SC phase, neutralized
by $\mu_8$. In particular it is no longer energetically favored.
 
Coming back to our original problem, the situation is quite similar:
When we apply the axial color transformation, \Eq{eq:axcol}, to the
color non-neutral 2SC phase, the $n_8$ color charge becomes reduced.
However, unlike in the case discussed above, it does not completely
reappear in other components $n_a$.  Instead, the transformation gives
rise to a non vanishing {\it axial} color density, \beq n^{(5)}_{a} =
\av{q^\dagger \gamma_5 \lambda_a q}\,.  \eeq To be precise, one finds
\begin{alignat}{2}
    &n_8\big|_{\text{p2SC}}&
    &= \frac{1}{4}(1 + 3 \cos\theta)\,n_8\big|_{\text{2SC}}\,,  
\nonumber\\
    &n_3\big|_{\text{p2SC}} &
    &= \frac{\sqrt{3}}{4}(1 - \cos\theta)\,n_8\big|_{\text{2SC}}\,,
\nonumber\\
    &\left.n_6^{(5)}\right|_{\text{p2SC}}&
    &= -\frac{\sqrt{3}}{2}\sin\theta\,n_8\big|_{\text{2SC}}\,.
\end{alignat}
Since the sum $\sum_a (n_a^2 + n_a^{(5)}\,^2)$ is conserved in the
transformation, the color charge remaining in the vector components
gets reduced.  Hence, since only the vector color densities $n_a$ are
required to vanish, the rotated solution can be neutralized more
easily.  This favors the p2SC phase, whereas on the other hand it is
disfavored by the nonvanishing up and down quark masses, which break
the spurious axial color symmetry explicitly.  The fact that the p2SC
phase only appears in a certain region of the phase diagram could thus
be explained by the interplay of these opposite effects.

\begin{figure}
  \includegraphics[width=0.47\textwidth]{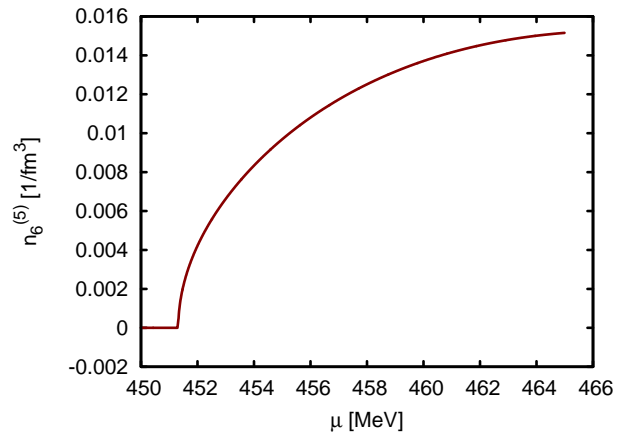}
  \caption{The axial color density $n_6^{(5)}$ for $H = 3/4\;G$ at 
    $T = 25$~MeV as a function of $\mu$. The values are nonzero in the 
    p2SC phase and smoothly go to zero at the second-order phase transition
    to the 2SC phase at $\mu = 451.3$~MeV.
  } 
  \label{n56}
\end{figure}

As shown in \Fig{n56}, we indeed find a nonvanishing $n_6^{(5)}$ in
the p2SC phase.  Since in QCD there are no ``axial gluons'', which
could couple to these axial color charges, there is in principle no
need to remove them by introducing corresponding ``axial color
chemical potentials''.  In this sense the axial color transformation
would be a legitimate way to reduce the free energy of the system.  On
the other hand, as we have seen before, it is only an accidental
(approximate) symmetry of the mean-field thermodynamic potential.  It
is therefore unlikely that this mechanism works beyond mean field,
i.e., the p2SC phase is most likely a model artifact.

Finally, we note that our numerical \cflk solutions have a small axial
color charge density, too.  Although, unlike in the p2SC phase, the
Goldstone modes which condense in the \cflk phase have a well-founded
physical basis, this could be a model artifact as well.  As already
noticed in \Ref{Buballa:2004sx}, the solutions of the gap equations
for the scalar and pseudoscalar diquark condensates in the \cflk phase
are more similar to an axial color rotation than to an axial flavor
rotation of the CFL solutions.  On the other hand, both
transformations lead to the same result in the chiral limit, i.e., the
difference is a higher-order correction in the quark mass.  Hence,
even if the preference of the axial color rotation was again a model
artifact, this would be only a small effect on top of the real
physical origin of the \cflk phase.


\section{Summary and conclusions}
\label{sec:conclusions}

In this paper we have reanalyzed the phase structure of homogeneous
electrically and color neutral quark matter within an NJL-type model.
Both, quark-antiquark condensates and diquark condensates have been
taken into account in order to describe dynamical mass generation and
color superconductivity on the same footing.  The main focus was on
the role of pseudoscalar diquark condensates, which allow for the
possibility of kaon condensation in the CFL phase.

A similar investigation was already performed some time ago 
in  Ref.~\cite{Warringa:2006dk}. 
We confirm the main result that the \cflk phase replaces major parts 
of the CFL phase in the phenomenologically interesting regime. 
Numerically, we also confirm that there are regions where a so-called
p2SC phase is favored, i.e., a two-flavor superconducting phase with
pseudoscalar condensates. We have argued, however, that this is most
likely a model artifact. 

Our main result is the existence of a gapped \cflk phase at small
enough chemical potential and very low temperature, $T \ll 10$~keV.
This is in contrast to Ref.~\cite{Warringa:2006dk}, where it was
claimed that the entire \cflk phase is completely gapped.  The
smallness of the critical temperature of the gapless region is due to
the fact that the mode contains quarks with nonvanishing $\tilde Q$
charge.  Like in the case of the $bu-rs$ mode in the gCFL
phase~\cite{Alford:2004hz}, this keeps the system close to the
unpairing line and, even at $T=0$, the blocking region stays very
small.  A small increase of the temperature is then sufficient to
restore the gap by Fermi smearing.

At $T=0$, our results are in qualitative agreement with effective
field theory, which predicts that the first gapless mode the \cflk
phase is
charged~\cite{Kryjevski:2004jw,Kryjevski:2004kt,Gerhold:2006np}.
However, as also observed in \Ref{Forbes:2004ww}, there are
quantitative differences.  For large coupling constants we even find
that the \cflk phase gets gapless at larger chemical potentials than
the CFL phase.  We also identified \cflk solutions with neutral
gapless modes, which have been predicted within effective field theory
as well~\cite{Kryjevski:2004jw,Kryjevski:2004kt,Gerhold:2006np}.  In
our analysis, however, these solutions were never favored.

It is well known that the gapless 2SC and CFL phases at low
temperatures suffer from chromomagnetic instabilities, signalling that
these phases are not the true ground state.  For the g\cflk phase with
charged gapless modes it was shown in \Ref{Gerhold:2006np} that all
Meissner masses are real.  Nevertheless the authors found a weak
instability with respect to the spontaneous formation of a
Goldstone-boson current, corresponding to a $p$-wave kaon
condensate~\cite{Kryjevski:2005qq,Schafer:2005ym}.  This, and the fact
that, at least for some parameters, we still found regions where the
gCFL phase is favored lead us to the conclusion that the phases
indicated in \Figs{phasediagram} and \ref{phasediagramHG} do not
always represent the true ground state of the system.  Most promising
candidates for improvements are inhomogeneous or anisotropic
phases. This includes the above-mentioned $p$-wave kaon condensates as
well a crystalline color superconducting
phases~\cite{Alford:2000ze,Bowers:2002xr,Casalbuoni:2003wh,GiannakisRen,
  Casalbuoni:2005zp,Mannarelli:2006fy,Rajagopal:2006ig,Nickel:2008ng,
  Sedrakian:2009kb}.


\section*{Acknowledgments}
We thank H.~Warringa for useful discussions and M.~Alford,
C.~Kouvaris, V.~Kleinhaus, K.~Rajagopal, and F.~Sandin for valuable
information about their numerical gCFL solutions.  The work of
H.B. was supported by the Helmholtz International Center for FAIR and
by the Helmholtz Graduate School for Hadron and Ion Research.
M.B. acknowledges partial support by EMMI.

\appendix

\section{Axial color symmetry of the mean-field thermodynamic potential
in the (p)2SC phase}
\label{app:p2SCsym}

In this appendix we show that, for $M_u = M_d = 0$ and vanishing color
chemical potentials, the mean-field thermodynamic potential,
\Eq{eq:OmegaMF}, is invariant under the axial color transformation,
\Eq{eq:axcol}, if the transformation is restricted to the non-strange
sector and only the diquark condensates $\Delta_{22}^{\scalar}$ and
$\Delta_{25}^{\pscalar}$ are present.

If $\Delta_{22}^\scalar$ and $\Delta_{25}^{\pscalar}$ are the only
non-vanishing diquark condensates, the inverse propagator $S^{-1}$,
\Eq{eq:Sinv}, can be put in a block-diagonal form with four $6\times
6$ and six $2\times 2$ blocks.  The $6\times 6$ blocks contain the
paired up and down quarks.  Since all colors participate in the
pairing, these blocks are larger than the corresponding blocks in the
2SC phase, which are $4\times4$ blocks.  The $2\times 2$ blocks
correspond to the (unpaired) strange quarks.  Note that for $M_u = M_d
= 0$ and, hence, $\phi_u = \phi_d = 0$, the strange sector completely
decouples from the non-strange sector in the (p)2SC phase. Therefore,
since we have restricted the axial color transformation to the
non-strange sector, the $2\times 2$ blocks will not be affected by the
transformation. Therefore we focus only on the $6\times 6$ blocks.

Without loss of generality, we may choose the three-momentum of the
quark to be parallel to the $z$-axis, i.e., $\vec p = p \vec e_3$.
For $M_u = M_d = 0$ and vanishing color chemical potentials the
$6\times 6$ matrices then take the form
\begin{widetext}
\begin{equation}\left(S_{u_b - d_r -
        u_g}^{\pm}\right)^{-1} =
    \left(\begin{array}{cccccc}p_0\pm\mu_u&-p&-\Delta_{25}^{\pscalar}&0&0&0
\\
-p&p_0 \pm\mu_u&0&\Delta_{25}^\pscalar&0&0
\\
-\Delta_{25}^\pscalar&0&p_0\mp\mu_d&p&0&\mp\Delta_{22}^\scalar
\\
0&\Delta_{25}^\pscalar&-p&p_0\mp\mu_d&\pm\Delta_{22}^\scalar&0
\\
0&0&0&\pm\Delta_{22}^\scalar&p_0\pm\mu_u&-p
\\
0&0&\mp\Delta_{22}^\scalar&0&-p&p_0\pm\mu_u
\\
\end{array}\right)
\label{p2SCblock}
\end{equation} 
\end{widetext}
and
\begin{equation} 
    \left(S^\pm_{d_b-u_r - d_g}\right)^{-1} = \left[ \left(S_{u_b
        - d_r -     u_g}^{\pm}\right)^{-1} \text{ with } u \leftrightarrow
    d\right].\label{p2SCblock2}
\end{equation}
Here $\mu_u$ ($\mu_d$) is the chemical potential for the up (down) quarks.
Taking the determinant, we obtain
\begin{alignat}{1}
      &\det\left[\left(S_{u_b - d_r -
          u_g}^{\pm}\right)^{-1}\right]
\nonumber\\
     &= \phantom{\times}\left[\left(p_0 \pm \mu_u\right)^2 +
      p^2\right]
\nonumber\\
      &\phantom{=}\times \left[\left(\left(p_0 \mp \mu_d\right) \pm
        p\right) \left(\left( p_0 \pm      \mu_u\right) \mp
        p\right)
       - (\Delta_{22}^{\scalar})^2 - (\Delta_{25}^{\pscalar})^2 \right]
\nonumber\\    
      &\phantom{=}\times \left[\left(\left(p_0\mp \mu_d\right) \mp p
       \right)\left(\left(p_0 \pm      \mu_u\right) \pm
        p\right)
       - (\Delta_{22}^{\scalar})^2 - (\Delta_{25}^{\pscalar})^2 \right]
\label{p2SCdet}
\end{alignat}
and the analogous result for 
$\det[(S^\pm_{d_b-u_r - d_g})^{-1}]$
with $\mu_u$ and $\mu_d$ interchanged.

We see that the determinants and, hence, $\Tr\ln S^{-1} = \ln\det
S^{-1}$ depend only on the sum $(\Delta_{22}^{\scalar})^2 +
(\Delta_{25}^{\pscalar})^2$, which is invariant under the axial color
rotation, cf.~\Eq{eq:condaxrot}.  The same is obviously true for
$\mathcal{V}$, \Eq{eq:V}.  Consequently, the mean-field thermodynamic
potential, \Eq{eq:OmegaMF}, is invariant, too.

Note that it was important that we have restricted ourselves to the
limit $M_u = M_d = 0$. Otherwise $\Omega$ would not be invariant.  In
particular, we would then have to transform the condensates $\phi_u$
and $\phi_d$ as well.  For the same reason, we have restricted the
transformation to the non-strange sector.  Alternatively, we could
have considered the case where the strange quark mass vanishes,
too. However, this would be a rather poor approximation for the real
situation in the (p)2SC phase.


\end{document}